# THEORETICAL DESCRIPTION OF FERROELECTRIC AND PYROELECTRIC HYSTERESIS IN THE DISORDERED FERROELECTRIC-SEMICONDUCTOR FILMS

A. N. Morozovska[1],

[1]V.Lashkaryov Institute of Semiconductor Physics, National Academy of Science of Ukraine,

41, pr. Nauki, 03028 Kiev, Ukraine, e-mail: morozo@i.com.ua

E.A. Eliseev[2]

[2]Institute for Problems of Materials Science, National Academy of Science of Ukraine,

3, Krjijanovskogo, 03142 Kiev, Ukraine, e-mail: eliseev@i.com.ua

**Abstract**

We have modified Landau-Khalatnikov approach and shown that both the polar lattice and the screened charged defects determine the response of disordered ferroelectric-semiconductors. This system exhibits the spatially inhomogeneous switching under the external field while Landau-Khalatnikov model describes homogeneous switching with the sharp pyroelectric coefficient peak near the thermodynamic coercive field value. Our model gives more realistic pyroelectric hysteresis loop shape without any peaks near the coercive field and describes both qualitatively and quantitatively typical $Pb(Zr,Ti)O_3$ and $(Sr,Ba)Nb_2O_6$ films pyroelectric hysteresis loops.

## 1. INTRODUCTION

The main peculiarity of ferroelectric materials is hysteresis of their dielectric permittivity $\varepsilon$, displacement $D$ and pyroelectric coefficient $\gamma$ over electric field $E_0$ applied to the sample [1], [2]. Ferroelectric $D(E_0)$, dielectric $\varepsilon(E_0)$, and pyroelectric $\gamma(E_0)$ hysteresis loops in an inhomogeneous ferroelectric-semiconductor film have several characteristic features depicted in the Fig.1a.

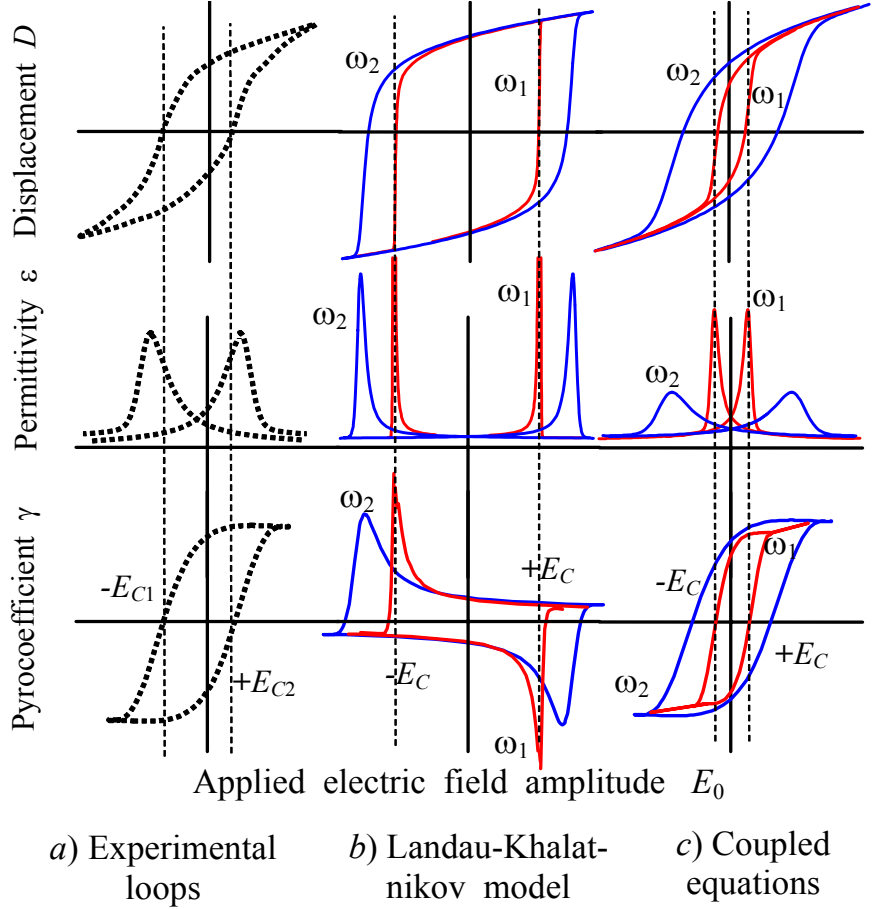

*a*) Experimental loops    *b*) Landau-Khalat-nikov model    *c*) Coupled equations

**Figure 1**. Dielectric $\varepsilon(E_0)$, ferroelectric $D(E_0)$ and pyroelectric $\gamma(E_0)$ hysteresis loops. Different plots correspond to the data obtained for a semiconductor-ferroelectric film (a), Landau-Khalatnikov model (b) and our coupled equations (c) for a bulk sample ($\omega_1 << \omega_2$ are two frequencies of applied electric field). Note, that Landau-Khalatnikov thermodynamic coercive field $E_C$ is at least several times greater than the experimentally observed values $E_{C1}$ and $E_{C2}$ (sometimes the loops observed in the films with thickness 50nm-5µm are shifted on the value $E_i = E_{C2} - E_{C1}$).

In particular ferroelectric hysteresis loops measured in PbZr$_x$Ti$_{1-x}$O$_3$ films doped with donor additives of Nb, Nd [3], [4], disordered and relaxor ferroelectrics films (e.g. Sr$_x$Ba$_{1-x}$Nb$_2$O$_6$ (SBN-x), PbZr$_{1-x}$Ti$_x$O$_3$: La (PZT: La), PbMg$_{1/3}$Nb$_{2/3}$O$_3$) have rather sloped and slim shape with low coercive fields (see e.g. [5], [6], [7], [8]), not the square one. Aforementioned ferroelectric hysteresis loops are obtained by conventional Sawyer-Tower method [2], [9], [10] at low frequencies of applied field changing. In these materials low frequency dielectric hysteresis loops were obtained from the measurements of the capacity dependence on the applied voltage [11], [12], [13]. The loops of pyroelectric hysteresis are usually obtained by means of pyroelectric response measurements under the periodic modulation of thermal flux when applied electric field varies independently with much lower frequency [13]. Pyroelectric loops usually are sloped and slim; no maximums of pyroelectric response near the coercive field are observed [13], [14].

Nowadays such typical ferroelectric-semiconductors as BaTiO$_3$ ceramics, disordered and relaxor Sr$_x$Ba$_{1-x}$Nb$_2$O$_6$ (SBN-x), PbZr$_x$Ti$_{1-x}$O$_3$ (PZT(x/1-x)) solid solutions slightly doped with Fe, Cr or La, Nb, Nd, Ce *etc.*, their films, multilayers and heterostructures are widely used in sensors, actuators, electro-optic, piezoelectric, pyroelectric devices and memory elements [2], [9], [10]. However, the task how to create ferroelectric material with pre-determined dielectric and/or pyroelectric properties is solved mainly empirically. The correct theoretical consideration and modeling of related problems seems rather useful both for science and applications. It could answer fundamental questions about the nature of lattice-defects correlations, possible self-organization in the system and help to tailor new ferroelectric-semiconductor materials.

Conventional phenomenological approaches [15], [16] with material parameters obtained from first-principle calculations [17], [18] can be successfully applied to the pure ferroelectrics materials [19], [20], but they give significantly incomplete picture of the dielectric, ferroelectric and pyroelectric hysteresis in the doped or inhomogeneous ferroelectrics-semiconductors (compare Fig.1b with Fig.1a).

In particular, pioneer Landau-Khalatnikov approach, evolved for the single domain pure ferroelectrics-dielectrics [21], [22], describes homogeneous switching but presents neither domain pinning nor domain nucleation, nor domain movement. The calculated values of thermodynamic coercive field [1] are significantly larger than its experimental values for real ferroelectrics [9], [23]. Observed hysteresis loops usually look much thinner and sloped than Landau-Khalatnikov ones [2], [7], [9], [24], [25], [26]. Also equations of Landau-Khalatnikov type describe homogeneous pyroelectric coefficient switching with the sharp peak near the thermodynamic coercive field value. We could not find any experiment, in which the pyroelectric coefficient peak near the coercive field had been observed. Usually the pyroelectric hysteresis loops of doped ferroelectrics have typical slim shape with coercive field values much lower than the thermodynamic ones. Therefore, the modification of Landau-Khalatnikov approach for inhomogeneous ferroelectrics-semiconductors seems necessary [27].

In our recent papers [28], [29], [30], [31], [32] we have considered the displacement fluctuations and extrinsic conductivity caused by charged defects and thus modified the Landau-Khalatnikov approach for the inhomogeneous ferroelectrics-semiconductors. The derived system of three coupled equations gives the correct description of dielectric and ferroelectric hysteresis loops shape and experimentally observed coercive field values (see Fig.1c). In this paper we evolve the proposed model [29] for pyroelectric response and demonstrate that the pyroelectric hysteresis loops of ferroelectrics-semiconductor films with charged defects could be successfully described by using six coupled equations.

## 2. THEORETICAL DESCRIPTION

### 2.1 The model and basic assumptions

The proposed model has been evolved under the following **assumptions**.

- We consider $n$-type ferroelectric-semiconductor film with sluggish randomly distributed impurity centers or/and growth defects. The charge density of defects $\rho_S(\mathbf{r})$ is characterized by the positive average value $\overline{\rho}_S$ and random fluctuations $\delta\rho_S(\mathbf{r})$, i.e. $\rho_S(\mathbf{r}) = \overline{\rho}_S + \delta\rho_S(\mathbf{r})$ (see Fig.2a).

- Movable screening charges distribution $\delta n(\mathbf{r},t)$ with Debye screening radius $R_D$ surround each charged center (see Fig.2b), so free carriers charge density $n(\mathbf{r},t) = \overline{n} + \delta n(\mathbf{r},t)$ is characterized by the negative average value $\overline{n}$ and modulation $\delta n(\mathbf{r},t)$.

- Hereinafter the dash designates the averaging of the function $f$ over the film volume $V$: $f = \overline{f} + \delta f(\mathbf{r},t)$. By definition $\overline{f}(t) = \dfrac{1}{V}\int\limits_V f(\mathbf{r},t)d\mathbf{r}$ and $\overline{\delta f(\mathbf{r},t)} = 0$, where $f = \{n, \rho_S, \mathbf{E}, \mathbf{D}, ...\}$. The charged defects distribution is quasi-homogeneous (i.e. $\overline{\delta\rho_S^2} = const$, $\overline{(\delta\rho_S)}_{x,y} = 0$ for every $z$ inside the film $-l/2 \le z \le l/2$). The average distance between defects is $d = a/\sqrt[3]{N_d}$ ($a$ is the lattice constant, $0 < N_d << 1$ is defect concentration). For a μm-thick film with $0.1\% \le N_d \le 10\%$ one estimate that $d << l$ (see Fig.2a).

- The film occupies the region $-l/2 \le z \le l/2$. We suppose that the potentials of electrodes are given, so that the external field $E_0(t)$ is applied along the polar axis $z$. Screening charges distribution is deformed in the external field $E_0(t)$, and the system "defect center $\delta\rho_S$ + screening cloud $\delta n$" causes electric field, current and displacement fluctuations.

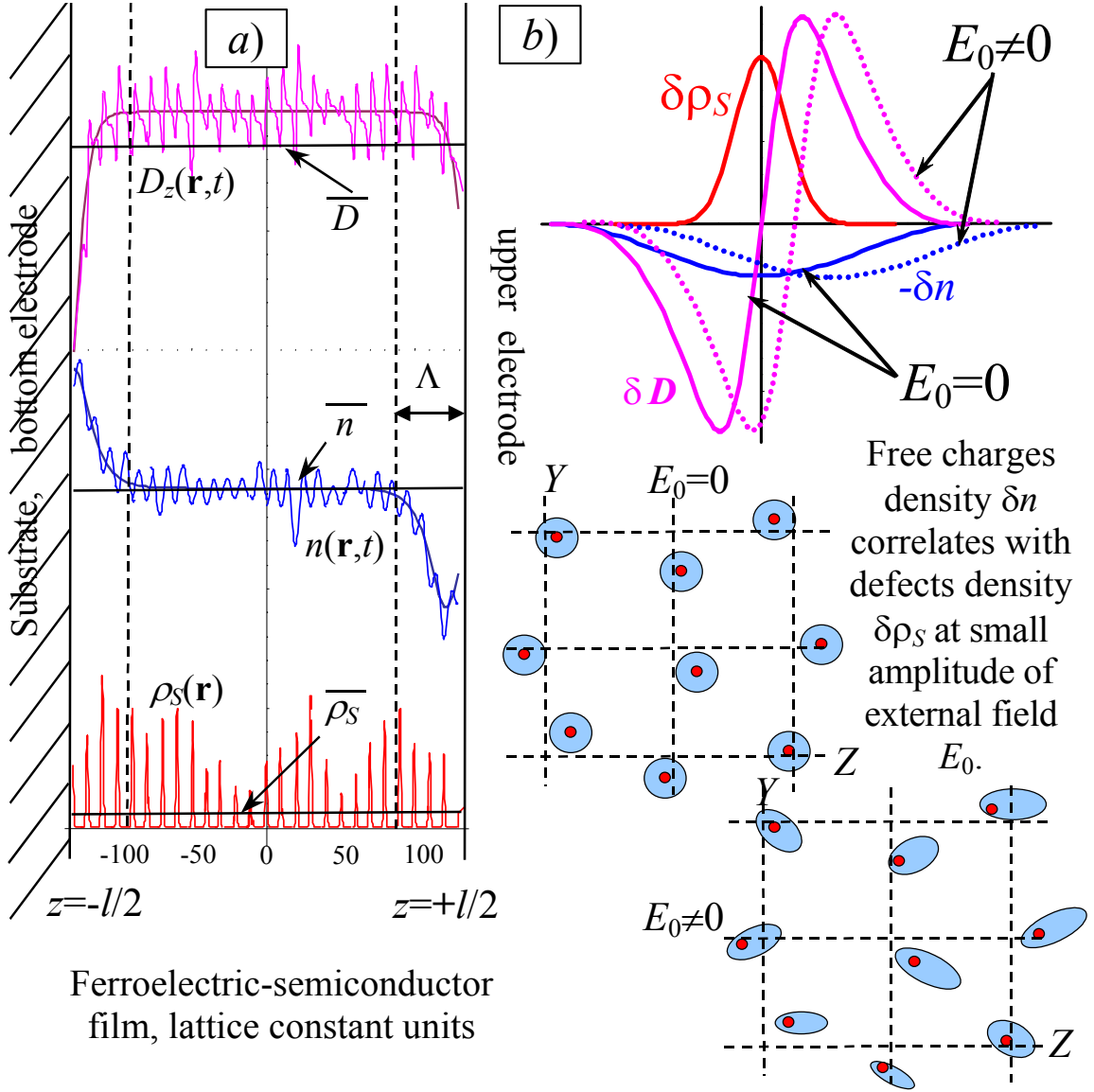

**Figure 2**. Spatial distribution of displacement $D$, free carriers with charge density $n$ and sluggish defects with charged density $\rho_S$ in an inhomogeneous ferroelectric-semiconductor film.

- The defects sluggishness means that $\overline{\left|\dfrac{\partial \delta \rho_S(\mathbf{r})}{\partial t}\right|} << \overline{\left|\dfrac{\partial \delta n(\mathbf{r},t)}{\partial t}\right|}$ and $\overline{\left|\dfrac{\partial \delta \rho_S(\mathbf{r})}{\partial t}\right|} << \overline{\left|\dfrac{dE_0}{l\,d\,t}\right|}$.

- The period of external field changing is small enough for the validity of the quasi-static approximation $rot\,\mathbf{E} = 0$, $rot\,\mathbf{H} = 0$. Thus, Maxwell's equations for the quasi-static inner electric field $\mathbf{E}(\mathbf{r},t)$ and displacement $\mathbf{D}(\mathbf{r},t)$ have the form:

$$div\,\mathbf{D} = 4\pi\big(\rho_S(\mathbf{r}) + n(\mathbf{r},t)\big), \qquad \frac{\partial \mathbf{D}}{\partial t} + 4\pi\,\mathbf{j}_c = 0 \qquad . \tag{1}$$

Here $\mathbf{E}(\mathbf{r},t) \equiv \mathbf{e}_z E_0(t) + \delta\mathbf{E}(\mathbf{r},t)$ and $\mathbf{D}(\mathbf{r},t) \equiv \mathbf{e}_z \overline{D(t)} + \delta\mathbf{D}(\mathbf{r},t)$. Maxwell's equations (1) have to be supplemented by the material expression for macroscopic free-carriers current charge density $\mathbf{j}_c \approx \mu \overline{n} \mathbf{E} - \kappa \, grad \, n$. One should take into account only electron mobility $\mu$ and diffusion coefficient $\kappa$, allowing for the defects sluggishness. Hereinafter $\overline{n} < 0$, $\mu < 0$ and $\kappa > 0$.

- We use the Landau-Khalatnikov equation for the displacement z-component relaxation as the equation of state, but take into account the charged defects and correlation effects (i.e. $\partial^2 D_z / \partial \mathbf{r}^2 \neq 0$, because $D_z(\mathbf{r}, t) = \overline{D(t)} + \delta D(\mathbf{r}, t)$):

$$\Gamma \frac{\partial D_z}{\partial t} + \alpha D_z + \beta D_z^3 - \lambda \frac{\partial^2 D_z}{\partial \mathbf{r}^2} = E_z \qquad (2)$$

Parameter $\Gamma > 0$ is the kinetic coefficient, $\alpha < 0$, $\beta > 0$ are material parameters, $\lambda > 0$ in the gradient coefficient of the hypothetical pure (free of defects) uniaxial ferroelectric. For perovskites one can use coupled Landau-Khalatnikov equations for three components $D_{z,x,y}(\mathbf{r}, t)$ instead of one equation (2) (see e.g. free energy in [9]). Hereinafter we suppose that only the coefficient $\alpha = -\alpha_T (T_C - T)$ essentially depend on temperature in usual form: $\partial \alpha / \partial T = \alpha_T$ [1]. The temperature dependence of other coefficients $\beta$, $\lambda$ and $\Gamma$ could be neglected within the framework Landau-Ginsburg phenomenology for the second order phase transitions [1]. The temperature dependence $\partial \overline{n} / \partial T$ could be neglected at temperatures $T >> E_a / k_B \sim 50\,K$ [33], [34].

The spatial distribution and temporal evolution of the displacement $\mathbf{D}(\mathbf{r}, t)$ in the film is determined by the non-linear system (1)-(2) supplemented by the initial distributions of all fluctuating variables (e.g. definite distribution of charged defects) and boundary conditions. In the Appendix A we show, that the system (1)-(2) is complete, because the quantities $\delta n$, $\delta \mathbf{E}$ can be expressed via the fluctuations of displacement $\delta D$ and $\delta \rho_s(\mathbf{r})$. The study of the domain wall pinning by the given distribution of charged defects, domain nucleation during spontaneous displacement reversal is beyond the scope of this paper. Therefore, hereinafter we consider only the averaged quantities. In the general case hypothetical complete exact system for average values and all their correlations could contain infinite number of differential equations and thus is not the simpler problem than the numerical solution of the stochastic differential equations [35]. Allowing for the aforementioned approximations and assumptions we have no special reason to consider an exact system. Instead we propose to deal with much simpler incomplete system for the average measurable quantities $\overline{D}$, $\overline{\delta D^2}$, *etc* and to estimate/postulate some correlations, namely:

- At external electric field $E_0(t)$ small in comparison with microscopic one we assume that

$$-\overline{\delta \rho_S \delta n(t)} \approx \eta(t) \overline{\delta \rho_S^2} \quad (0 \leq \eta(t, E_0) < 1), \qquad (3)$$

because each charged region $\delta\rho_S$ is surrounded by the electron density $\delta n$ in accordance with Debye screening mechanism, but the smooth part of charge density $\delta n$ accommodated near the surfaces in the layer with thickness $\Lambda$ and related to the spontaneous displacement screening does not correlate with random charged defects distribution $\delta\rho_S$ (see Figs 2b, Appendix A and [29] for details).

- Taking into account that under the conditions $d << l$ and $d << \Lambda$ the frequency spectrum of fluctuations $\delta D$ and $\delta\rho_S$ has well localized maximum at spatial frequency $2\pi/d$, one obtains that:

$$\overline{\delta D \frac{\partial^2 \delta D}{\partial \mathbf{r}^2}} \approx -\frac{1}{d^2}\overline{\delta D^2}, \quad \overline{\delta\rho_S \frac{\partial^2 \delta D}{\partial \mathbf{r}^2}} \approx -\frac{1}{d^2}\overline{\delta\rho_S \delta D} \quad (4)$$

See Appendix A, Fig. 2a and Fig. 3 for illustration in one-dimensional case.

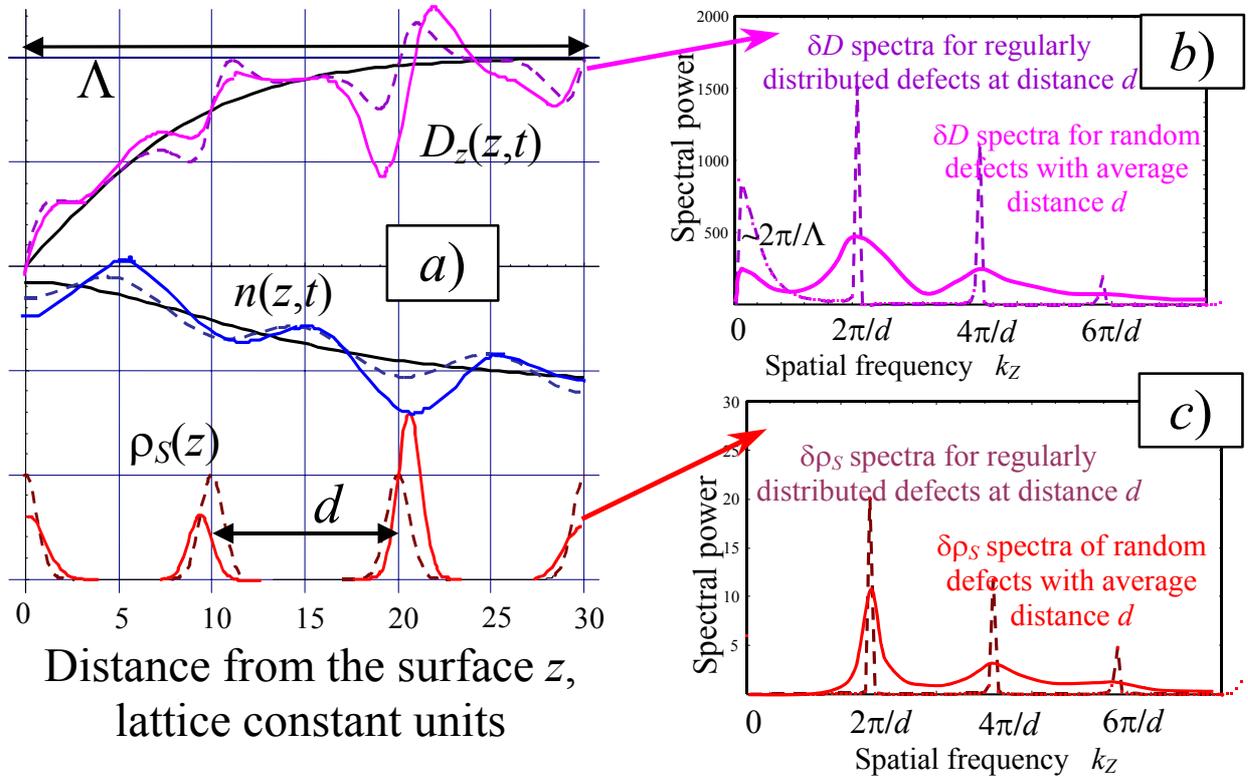

**Figure 3**. 1D-spatial distribution and frequency spectrum of $\delta D(z,t)$ and $\delta\rho_S(z)$ calculated for average distance between defects $d = 10$, screening layer thickness $\Lambda = 30$, film thickness $l \approx 260$ lattice constants and charged defects linear concentration $\sim 10\%$. Dashed peaks correspond to the model 1D-spectrum when identical defects are regularly distributed with distance $d$ between them (see dashed curves). Solid curves are computer simulation of the spectrum with random defects with average distance $d$ between them. Note, that $\delta D(z,t)$ low frequency peak with halfwidth $\sim 2\pi/\Lambda$ corresponds to the smooth displacement variation in the screening layer with thickness $\Lambda$.

## 2.2 Coupled equations for average quantities

In our papers [29], [30], [31], we modify classical Landau-Khalatnikov equation (3) for bulk samples and obtain the system of coupled equations for average displacement $\overline{D}$, its mean-square fluctuation $\overline{\delta D^2}$ and correlation with charge defects density fluctuations $\overline{\delta D \delta \rho_S}$. In the present paper we evolve coupled equations for a finite film with thickness $l$ and quasi-homogeneous distribution of charged defects (see Appendix A for details).

The average pyroelectric coefficient $\overline{\gamma}$ can be easily obtained by the differentiation of the coupled system [29] on the temperature $T$. In this way we obtained coupled equations for the average displacement $\overline{D}$, its mean-square fluctuation $\overline{\delta D^2}$ and correlation $\overline{\delta D \delta \rho_S}$, pyroelectric coefficient $\overline{\gamma} = \partial \overline{D} / \partial T$, its correlations with displacement fluctuations $\overline{\delta \gamma \delta D} = \overline{\delta D \, \partial \delta D / \partial T}$ and charge defects density fluctuations $\overline{\delta \gamma \delta \rho_S} = \partial \overline{\delta D \delta \rho_S} / \partial T$ :

$$\Gamma \frac{\partial \overline{D}}{\partial t} + \left(\alpha + 3\beta \overline{\delta D^2}\right)\overline{D} + \beta \overline{D}^3 = E_0(t) + E_i(l,t), \tag{5}$$

$$\frac{\Gamma_R}{2} \frac{\partial \overline{\delta D^2}}{\partial t} + \left(\alpha_R + 3\beta \overline{D}^2\right)\overline{\delta D^2} + \beta(\overline{\delta D^2})^2 = E_0(t)\left(\frac{\overline{\delta D \delta \rho_S}}{\overline{n}} - \delta E_i\right) + \frac{4\pi k_B T}{\overline{n} e} \overline{\delta \rho_S (\delta \rho_S + \delta n)} \ , \tag{6}$$

$$\Gamma_R \frac{\partial \overline{\delta D \delta \rho_S}}{\partial t} + \left(\alpha_R + 3\beta \overline{D}^2 + \beta \overline{\delta D^2}\right) \overline{\delta D \delta \rho_S} = -E_0(t)\frac{\overline{\delta \rho_S \delta n}}{\overline{n}} \ . \tag{7}$$

$$\Gamma \frac{\partial \overline{\gamma}}{\partial t} + \left(\alpha + 3\beta \overline{\delta D^2} + 3\beta \overline{D}^2\right)\overline{\gamma} = -\left(\alpha_T + 3\beta \overline{\delta \gamma^2}\right)\overline{D}, \tag{8}$$

$$\Gamma_R \frac{\partial \overline{\delta \gamma \delta D}}{\partial t} + 2\left(\alpha_R + 2\beta \overline{\delta D^2} + 3\beta \overline{D}^2\right)\overline{\delta \gamma \delta D} = \left(\begin{matrix} E_0(t)\dfrac{\overline{\delta \gamma \delta \rho_S}}{\overline{n}} - \left(\alpha_{RT} + 6\beta \overline{D}\,\overline{\gamma}\right)\overline{\delta D^2} + \\[2mm] + \dfrac{4\pi k_B}{\overline{n} e}\overline{\delta \rho_S(\delta \rho_S + \delta n)} \end{matrix}\right), \tag{9}$$

$$\Gamma_R \frac{\partial \overline{\delta \gamma \delta \rho_S}}{\partial t} + \left(\alpha_R + \beta \overline{\delta D^2} + 3\beta \overline{D}^2\right) \overline{\delta \gamma \delta \rho_S} = -\left(\alpha_{RT} + 2\beta \overline{\delta \gamma \delta D} + 6\beta \overline{D}\,\overline{\gamma}\right)\overline{\delta D \delta \rho_S} \ . \tag{10}$$

The built-in electric field $E_i(l,t) = \dfrac{4\pi\lambda}{l}\overline{(\delta n)}_{x,y}\Big|_{-l/2}^{+l/2}$ in (5) and its deviation

$\delta E_i = \dfrac{2\pi}{l\,\overline{n}}\overline{\left(\int\limits_{z_0}^{\tilde{z}} dz (\delta n + \delta \rho_S)\right)^2}_{x,y}\Bigg|_{-l/2}^{+l/2}$ in (6) are inversely proportional to the film thickness $l$, thus they

vanish in the bulk material [29]. For a finite film built-in field $E_i(l,t)$ is induced by the smooth part of space charges $\delta n$ accommodated near the non-equivalent boundaries $z = \pm l/2$ (e.g. near the substrate with bottom electrode and free surface with upper electrode depicted in Fig.3). Such layers are created by the screening carries [15], [18]. Built-in field value completely determines the horizontal shift of hysteresis loops observed in the ferroelectric films with thickness from dozen nanometers up to several microns [36], [37]. It is known that built-in field diffuses paraelectric-ferroelectric phase transition, in particular it shifts and smears dielectric permittivity temperature maximum [1]. Thus the film thickness decrease leads to the degradation of their ferroelectric and dielectric properties [27] up to the appearance of relaxor features [38].

Bratkovsky and Levanyuk [20] predicted the existence of built-in field in a finite ferroelectric film within the framework of phenomenological consideration. The field had a rather general nature. Our statistical approach confirms their assumption and gives the possible expression of the field existing in the inhomogeneous ferroelectric-semiconductor film. Another partial case, when the inner electric field is induced by the film-substrate misfit strain has been considered in our recent paper [39]. Misfit-induced field could be the main reason of the imprint only in the ultrathin strained films with the thickness less than 50nm.

The renormalization $\Gamma_R \equiv \Gamma + \tau_m$ of Khalatnikov kinetic coefficient in (6)-(7), (9)-(10) is connected with the free carrier relaxation with characteristic time $\tau_m = \dfrac{1}{8\pi\mu\overline{n}}$. The renormalization of coefficients $\alpha_R \equiv \alpha + \left(\lambda + k_B T / 4\pi\overline{n}e\right)\!/d^2$ and $\alpha_{RT} \equiv \alpha_T + k_B / 4\pi\overline{n}ed^2$ is caused by correlation and screening effects (see (4), Appendix and [30]). Coefficient $\alpha = -\alpha_T (T_C - T)$ is negative in the perfect ferroelectric phase without defects ($\overline{\delta\rho_S^2} = 0$, $T_C > T$). For the partially disordered ferroelectric with random defects ($\overline{\delta\rho_S^2} > 0$) coefficient $\alpha_R$ is positive, moreover $\alpha_R >> |\alpha|$, $\alpha_{RT} >> \alpha_T$ and usually the ratio $\xi = -\alpha_R/\alpha >> 1$ [29]. For example, for Pb(Zr,Ti)O$_3$ solid solution $\alpha \sim -(0.4 \div 2)\cdot 10^{-2}$ [9], gradient term $\lambda \approx 5\cdot 10^{-16} cm^2$, screening radius $R_D \sim \left(10^{-6} \div 10^{-4}\right) cm$ [15], average distance between defects $d \sim \left(10^{-6} \div 10^{-4}\right) cm$ and thus $\alpha_R \approx k_B T / 4\pi\overline{n}ed^2 \sim 10^2$. The source of fluctuations $\overline{\delta\rho_S(\delta\rho_S + \delta n)}$ in (6), (9) originates from the correlation between diffusion current and displacement fluctuations (see (A.19) in Appendix).

Let us introduce the following dimensionless variables in the system (5)-(10): $D_m = \overline{D}/D_S$ disorder parameter $\Delta_m = \overline{\delta D^2}/D_S^2$, $K_m = -\overline{\delta D \delta\rho_S}/D_S\overline{n}$, $\gamma_m = \overline{\gamma}/\gamma_S$, $\Pi_m = \overline{\delta\gamma\delta D}/D_S\gamma_S$ and $M_m = -\overline{\delta\gamma\delta\rho_S}/\gamma_S\overline{n}$, where $D_S(T) = \sqrt{-\alpha(T)/\beta}$ is bulk spontaneous displacement, $\gamma_S = dD_S(T)/dT$ is bulk pyroelectric coefficient. Note, that $\overline{n} = -\overline{\rho}_s$ allowing for the sample electro-neutrality. Dimensionless system (5)-(10) acquires the form:

$$\frac{\partial D_m}{\partial \tau} + \left(-1 + 3\Delta_m\right)D_m + \beta D_m^3 = E_e(\tau) + E_m(l)\,, \tag{11}$$

$$\frac{\tau_R}{2}\frac{\partial \Delta_m}{\partial \tau} + \left(\xi + 3D_m^2\right)\Delta_m + \Delta_m^2 = -E_e(\tau)\left(K_m + \delta E_m(l)\right) + gR^2\,, \tag{12}$$

$$\tau_R \frac{\partial K_m}{\partial \tau} + \left(\xi + 3D_m^2 + \Delta_m\right)K_m = -E_0(\tau)R^2\,, \tag{13}$$

$$\frac{\partial \gamma_m}{\partial \tau} + \left(-1 + 3\Delta_m + 3D_m^2\right)\gamma_m = -\left(1 + 3\Pi_m\right)D_m^2\,, \tag{14}$$

$$\tau_R \frac{\partial \Pi_m}{\partial \tau} + 2\left(\xi + 2\Delta_m + 3D_m^2\right)\Pi_m = \begin{pmatrix} -E_e(\tau)M_m - \left(\xi\left(\dfrac{T_C}{T}-1\right)+6\,D_m\,\gamma_m\right)\Delta_m + \\ + g\left(\dfrac{T_C}{T}-1\right)R^2 \end{pmatrix}\,, \tag{15}$$

$$\tau_R \frac{\partial M_m}{\partial \tau} + \left(\xi + \Delta_m + 3D_m^2\right)M_m = -\left(\xi\left(\frac{T_C}{T}-1\right)+6\,D_m\,\gamma_m + 2\Pi_m\right)K_m\,. \tag{16}$$

In accordance with our estimations [29], [30] fluctuations dimensionless amplitude $g = 4\pi k_B T \cdot \overline{n}\beta\left(\eta^{-1}-1\right)/\alpha^2 e$ varies in the range from $10^2$ to $10^4$ for typical ferroelectrics-semiconductors and $\eta(\tau, E_e) \sim \left(10^{-2}-1\right)$. In general case the correlator $R^2 = -\overline{\delta\rho_S \delta n(\tau)}/\overline{n}^2 = \eta(\tau)\overline{\delta\rho_S^2}/\overline{\rho}_S^2$ varies in the range $(0;\,1)$ because its amplitude is proportional to the charged defects disordering $\overline{\delta\rho_S^2}/\overline{\rho}_S^2$ (use (3) and sample electro-neutrality condition $\overline{n} = -\overline{\rho}_s$). The ratio $\xi = -\alpha_R/\alpha \approx \alpha_{RT}T/\alpha_T(T_C-T) \sim 10^2$, relaxation time $\tau_R = \Gamma_R/\Gamma \sim 1$, dimensionless time $\tau = -t/\alpha$. Also dimensionless external field $E_e = E_0/E_C$, built-in field $E_m = E_i/E_C$ and its deviation $\delta E_m = \delta E_i/E_C$ are introduced in (7)–(8) ($E_C = -\alpha\sqrt{-\alpha/\beta}$ is proportional to the coercive field of the perfect material with $\overline{\delta\rho_S^2} = 0$). Note, that $E_m$ is almost independent over disordering $R^2$ as caused by the smooth part of space charges $\delta n$, whereas $\delta E_m$ is proportional to the defects disordering $R^2$, namely $\delta E_m \sim \left(\eta^{-1}-1\right)R^2$ in our model (see Figs 2-3 and (3)).

Hereinafter we discuss only the system response near the equilibrium states, e.g. for the harmonic applied field $E_0(t) = E_a \sin(\omega t)$ the dimensionless frequency $w = -\Gamma\omega/\alpha$ should be smaller than unity. Additionally we assume that the series $\eta(\tau, E_e) = \eta_0 + \sum_{i=1}^{\infty} \eta_i(w, E_a) \sin(mw\tau)$ quickly converges and $\eta_0 >> |\eta_i|$, so we can substitute unknown functions $R^2(\tau)$, $g(\tau)$, $E_m(\tau)$, $\delta E_m(\tau)$ by their stationary parts $R_0^2$, $g_0$, $E_{m0}$, $\delta E_{m0}$ and then omit subscript "0" for the sake of simplicity. However we realize that it is only the first approximation, in fact correlation $\eta(\tau) \sim \overline{\delta\rho_S \delta n(\tau)}$ should slightly decrease with external field frequency or amplitude increase. This result could be explained allowing for the fact, that the dipole correlations "sluggish defect $\delta\rho_S$ + mobile carriers $\delta n$" weaken with the frequency or amplitude increase.

Thus the system (11)-(16) steady-state behavior is determined by the dimensionless built-in fields $E_m$, $\delta E_m$, frequency $w$ as well as by the aforementioned parameters $\xi$, $R^2$, $g$ and ratio $T/T_C$ It is appeared, that under the conditions $w < 1$, $g >> 1$ and $\xi >> 1$ the scaling parameter $gR^2/\xi$ determines the loops shape. Built-in field $E_m$ determines loops horizontal shift, whereas $\delta E_m$ determines their vertical asymmetry. However the visible vertical asymmetry exists only under unlikely condition $|\delta E_m| \geq 1$. The loops shape and position are almost independent over $\delta E_m$ under the condition $|\delta E_m| << 1$. The loops asymmetry becomes more pronounced with $R^2$ increasing. Within the framework of our model possible asymmetry is the main difference between the ferroelectric loops calculated for the film in comparison with the ones considered in [29] .for the bulk material.

Figs 4 demonstrate the typical changes of pyroelectric and ferroelectric hysteresis loop caused by the increase of charged defects disordering (note, that $gR^2/\xi \sim \overline{\delta\rho_S^2}/\overline{\rho_S^2}$). They prove, that disorder parameters $\overline{\delta\gamma^2} \sim \overline{\delta D^2}$ exists near the coercive field, where $\overline{D} \to 0$.

Let us underline, that in contrast to Landau-Khalatnikov equations describing homogeneous switching with $\overline{\delta D^2} \equiv 0$ near the coercive field, our coupled equations reveal **inhomogeneous** ferroelectric and pyroelectric response switching [29]. Namely, when the external field reaches the coercive value, the sample splits into the oppositely polarized regions, so that it is non-polarized as a whole. Every polarized region causes displacement and pyroelectric coefficient fluctuations $\delta\gamma$ and $\delta D$.

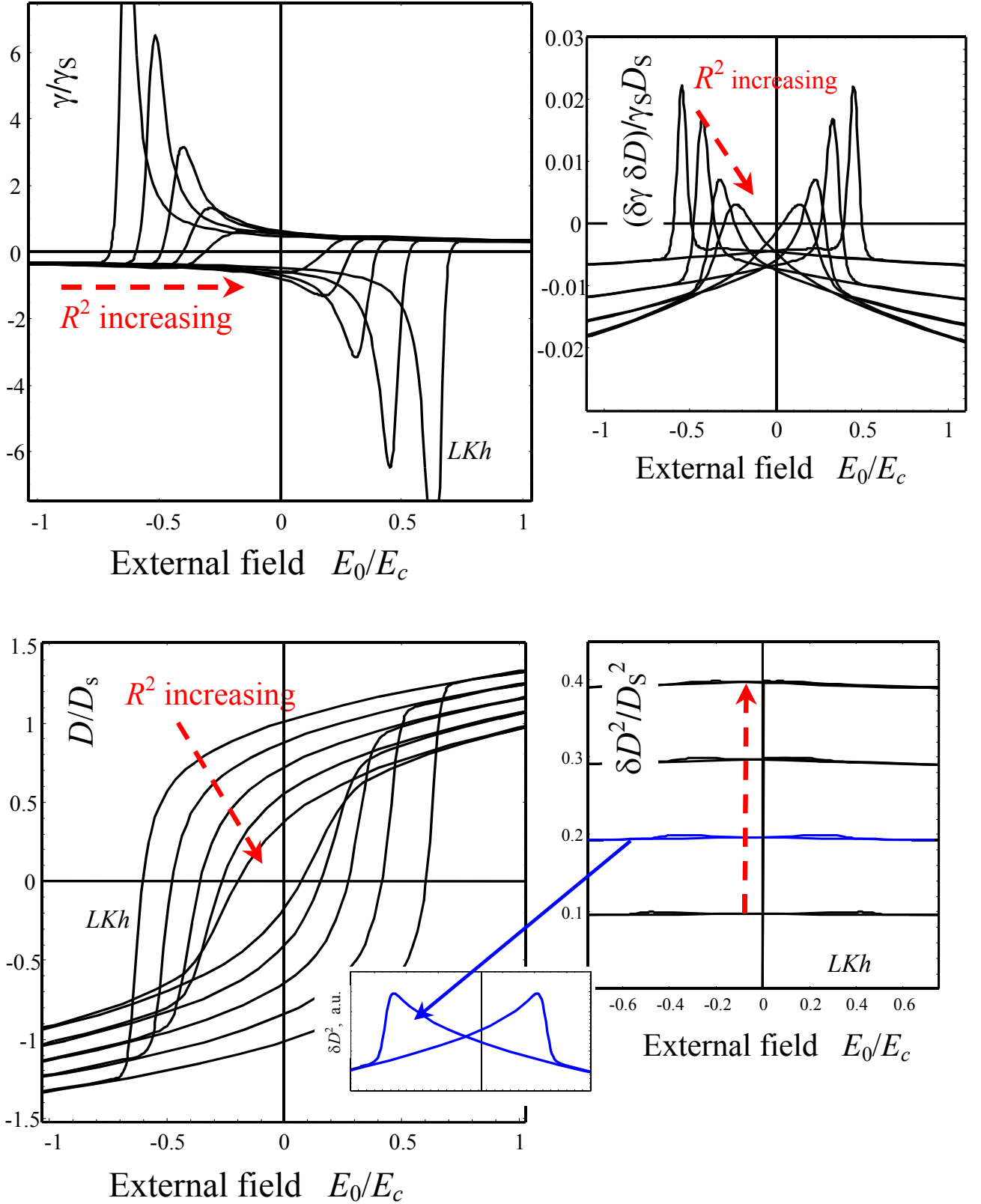

**Figure 4a.** Hysteresis loops of pyroelectric coefficient $\overline{\gamma}/\gamma_S$, its correlation $\overline{\delta\gamma\,\delta D}/(\gamma_S D_S)$, order parameter $\overline{D}/D_S$ and disorder parameter $\overline{\delta D^2}/D_S^2$ for small values $R^2 = 0;\,0.1;\,0.2;\,0.3;\,0.4$. Other parameters: $g$=100, $\xi$=100, $T/T_C$=0.45, $w$=0.025 and $E_m = 0.05$, $\delta E_m = 0.5 \cdot R^2$.

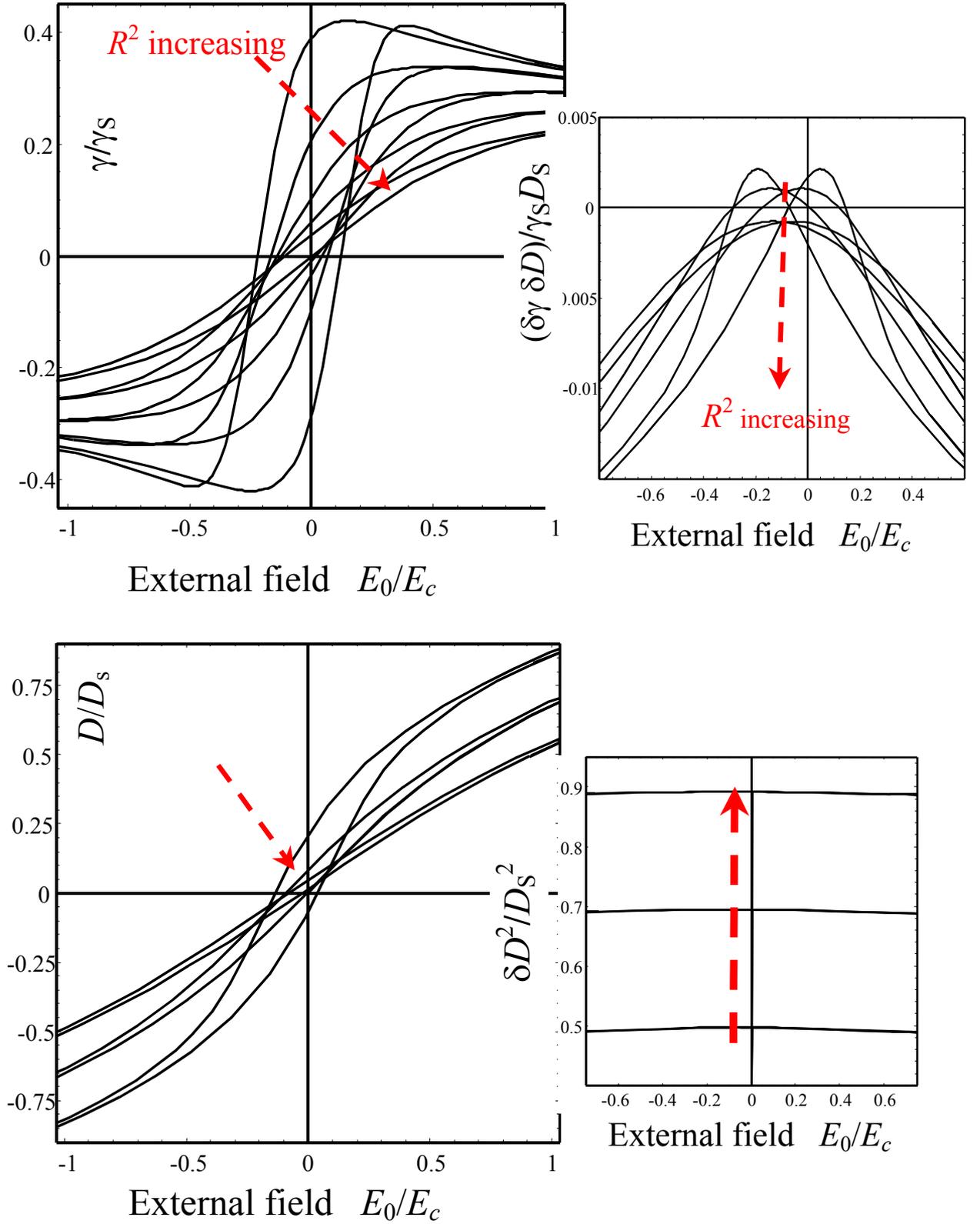

**Figure 4b.** Hysteresis loops of pyroelectric coefficient $\overline{\gamma}/\gamma_S$, its correlation $\overline{\delta\gamma\delta D}/(\gamma_S D_S)$, order parameter $\overline{D}/D_S$ and disorder parameter $\overline{\delta D^2}/D_S^2$ for higher values $R^2 = 0.5; 0.7; 0.9$. Other parameters: $g$=100, $\xi$=100, $T/T_C$=0.45, $w$=0.025 and $E_m = 0.05$, $\delta E_m = 0.5 \cdot R^2$.

It is clear from Figs 4a-b, that the increase of $gR^2/\xi$ value leads to the essential decrease and smearing of pyroelectric coefficient peaks near coercive field and to the decrease of the coercive field value (compare Landau-Khalatnikov loops with the other ones). At $R^2 \geq 0.5$ pyroelectric coefficient peaks near coercive field completely disappears and typical pyroelectric hysteresis loop looks like the ferroelectric one. At $R^2 \geq 0.7$ the coercive field is much smaller than its thermodynamic value at $R^2 = 0$. Thus pyroelectric and ferroelectric loops become sloped, much thinner and little lower under the increase charged defects disordering $\overline{\delta\rho_S^2}$. This effect is similar to the well-known **"square to slim transition"** of the ferroelectric hysteresis loops in relaxor ferroelectrics [24].

The changes of dielectric permittivity hysteresis caused by the increase of charged defects disordering in presented in Figs 5. It is clear that disordering $R^2$ increase shifts and smears dielectric permittivity maximum and thus leads to the degradation of the film ferroelectric and dielectric properties. Built-in field mainly shifts the hysteresis loops in the film.

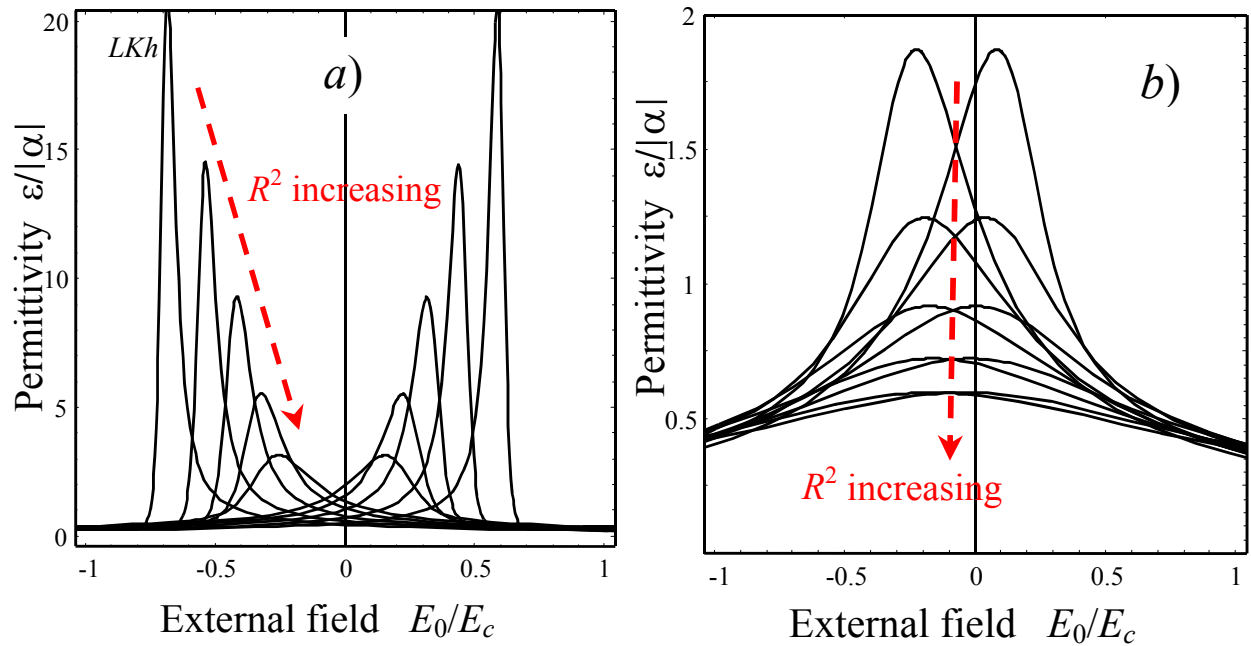

**Figure 5.** Hysteresis loops of dielectric permittivity $\overline{\varepsilon}/|\alpha|$ for different $R^2$ values: $R^2 = 0; 0.1; 0.2; 0.3; 0.4$ (a) and $R^2 = 0.5; 0.6; 0.7; 0.8; 0.9$ (b). Other parameters: $g$=100, $\xi$=100, $T/T_C$=0.45, $w$=0.025 and $E_m = 0.05$, $\delta E_m = 0.5 \cdot R^2$. Calculations were performed using (A.32-34) derived in Appendix.

Let us underline, that we do not know any experiment, in which pyroelectric coefficient peaks near coercive field had been observed. Moreover, usually pyroelectric hysteresis loops in doped ferroelectrics have typical "slim" shape with coercive field values much lower than the thermodynamic one [13], [14]. Though simplifications and approximations have been made (e.g. (3), (4)), our approach qualitatively explains available experimental results. The quantitative comparison of our numerical modeling with typical PZT and SBN-pyroelectric loops is presented in the next section.

## 2.3. Comparison with experimental results and discussion

Dopants, as well as numerous unavoidable oxygen $O^{-2}$ vacancies, can play a role of randomly distributed charged defects in "soft" PZT. Really, it is known that low level (0.001-1.0 %) donor additives La, Nb, Nd or Ce soften PZT dielectric and pyroelectric properties at room temperature. Ferroelectric and pyroelectric hysteresis loops have got relatively high $\gamma$ and $D$ remnant values, but reveal low coercive fields in soft PZT [9], [23]. Usually pyroelectric hysteresis loops of PZT are rather slim and sloped even at low frequencies $\omega \sim (0.1 \div 10) Hz$ [9], any pyroelectric coefficient maximum near the coercive field is absent [13], [14].

The pyroelectric response $U_\pi \sim \overline{\gamma}$ of the Pt-PZT-Pt/Ti-SiO$_2$/Si structure with oriented 2μm thick PZT(46/54): Nb film was obtained in [13] with the help of modulated thermal flux. During the measurements quasi-static applied voltage varied in the range (-11V, +11V) with low-frequency $\omega \sim 0.2 \ Hz$, the temperature $T$ changes with the frequency about 20 Hz. Ferroelectric and dielectric hysteresis loops measured for the same PZT films by using conventional Sawyer-Tower method and impedance analyzer at higher frequencies $\omega \sim 1 \ Hz$. Note, that a film capacity is proportional to the dielectric permittivity $\overline{\varepsilon}$ (see equations for $\overline{\varepsilon}$ in the end of Appendix). Stabilized loops (i.e. obtained after many circles) and our calculations are compared in the Fig.6.

It is clear from the Fig 6b, that ferroelectric loop is slightly asymmetrical, i.e. built-in electric field $E_i \neq 0$, whereas the pyroelectric loop measured at lower frequency look almost symmetrical. Note, that sometimes even stabilized loops are asymmetrical [39]. For instance even rather thick films of disordered ferroelectric SBN grown on different substrates typically reveal shifted ferroelectric and pyroelectric loops [14]. The pyroelectric loop of 6μm-thick SBN-0.5 film on Si-SiO$_2$ substrate is compared with our calculations in the Fig.7.

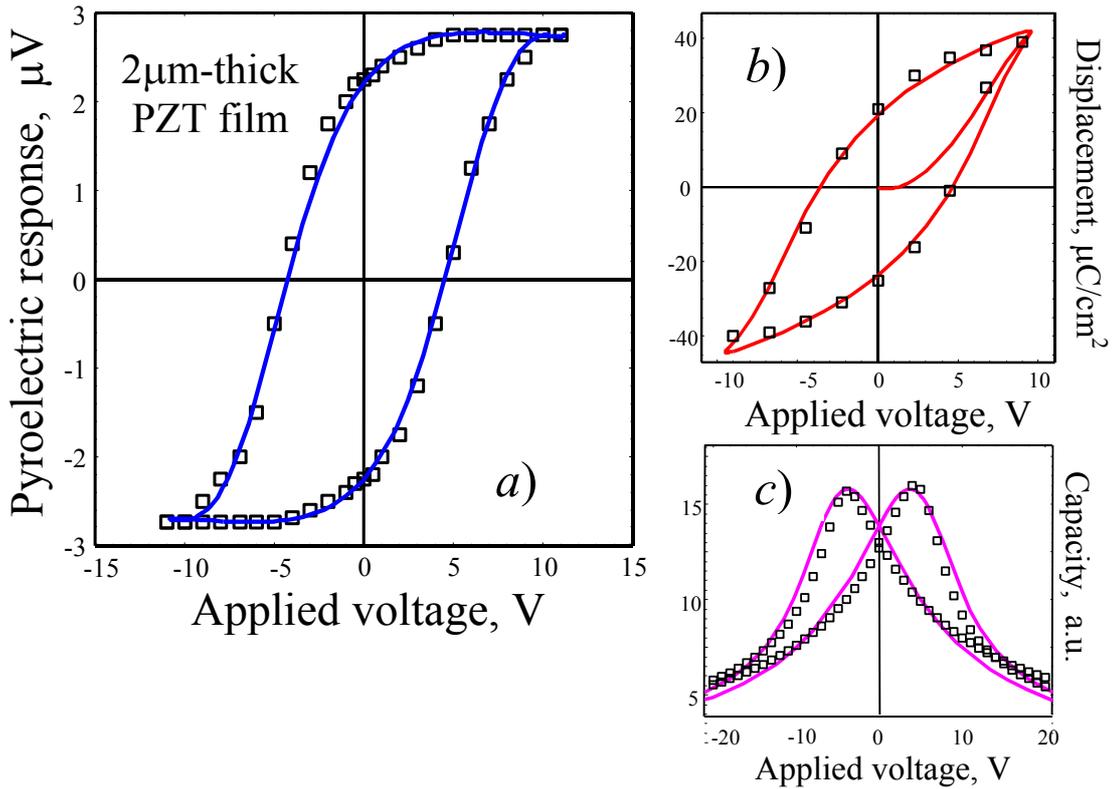

**Figure 6**. Typical pyroelectric (a) and ferroelectric (b) loops of 2μm-thick PZT (46/54): Nb film in on Ti-SiO₂/Si substrate. Squares are experimental data measured by Bravina *et al.* [13] at $T$=300 K, solid curves are our calculation with the fitting parameters $R^2$=0.5, $g$=100, $\xi$=100, $E_m$= -0.03 and $w$=0.11 (a), $w$=0.6 (b) and $w$=0.4 (c). All other parameters correspond to the ones for perfect PZT(46/54) solid solution [9].

It is clear from the Figs 6-7, that our model both qualitatively and quantitatively describes pyroelectric hysteresis loops in thick "soft" PZT and relaxor SBN films. Our modeling of ferroelectric and dielectric hysteresis loops was performed earlier (see e.g. [29]). Thus, despite used simplifications and approximations may require more background, verification and/or specification, the modeling based on the coupled equations (5)-(10) gives realistic coercive field values and typical pyroelectric hysteresis loop shape, in contrast to the Landau-Khalatnikov approach, that describes the homogeneous pyroelectric coefficient switching. Taking into account, that the inhomogeneous switching of spontaneous polarization and pyroelectric response occurs in the inhomogeneous and disordered ferroelectrics-semiconductors, the proposed coupled equations can be more relevant for the phenomenological description of their polar and pyroelectric properties, than the models based on Landau-Khalatnikov phenomenology.

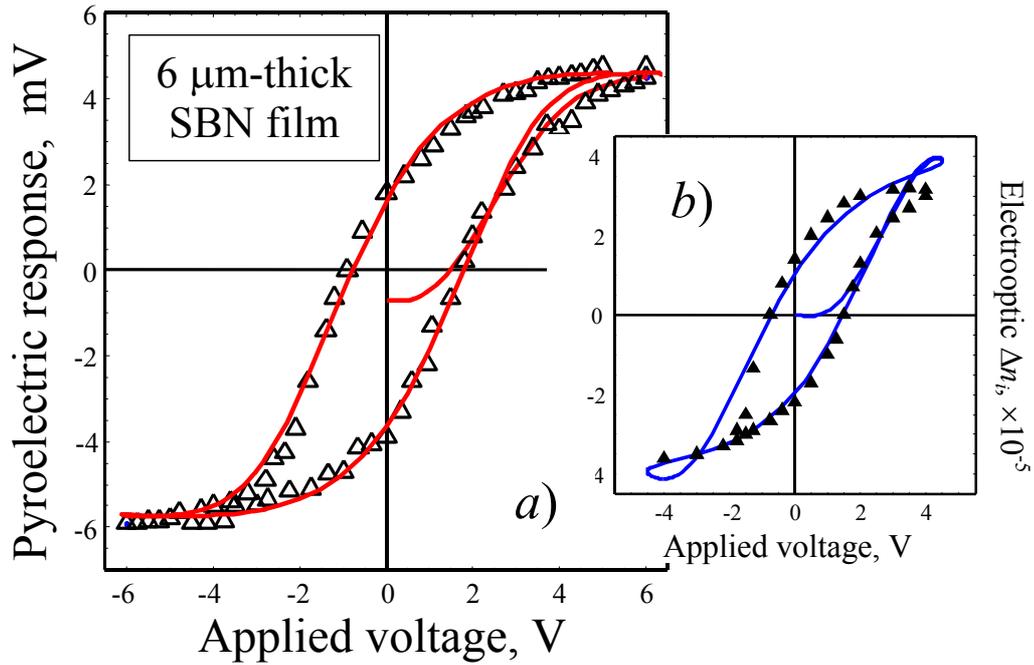

**Figure 7.** Pyroelectric response $U_\pi \sim \overline{\gamma}$ (a) and electro-optic modulation $\Delta n_i \sim \overline{\varepsilon} \cdot \overline{D}$ ((b) at $\omega \sim 1\,kHz$) obtained for 6μm-thick SBN-0.5 film on Si-SiO$_2$ substrate. Squares are experimental data measured by Kostsov *et al.* [14] at room temperature, solid curve is our calculation with the fitting parameters $R^2$=0.58, $g$=100, $\xi$=100, $E_m$= -0.013 and $w$=0.08 (a) and $w$=0.2 (b). All other parameters correspond to the ones for perfect SBN-0.5 [1].

## APPENDIX

We assume that the period of external field changing is small enough for the validity of the quasi-static approximation $rot\,\mathbf{E} = 0,\quad rot\,\mathbf{H} = 0$. Thus, Maxwell's equations for the quasi-static electric field $\mathbf{E}$ and displacement $\mathbf{D}$ have the form:

$$
\begin{aligned}
div\,\mathbf{D} &= 4\pi\big(\rho_S(\mathbf{r}) + n(\mathbf{r},t)\big), \\
\frac{\partial \mathbf{D}}{\partial t} &+ 4\pi\,\mathbf{j}_c = 0
\end{aligned}
\tag{A.1}
$$

They have to be supplemented by the material expression for current charge density:

$$
\mathbf{j}_c = \sum_{\rho_m = \rho_S,\,n}\big(\mu_m \rho_m \mathbf{E} - \kappa_m\,grad\,\rho_m\big) \approx \mu n\mathbf{E} - \kappa\,grad\,n
\tag{A.2}
$$

Here $\mathbf{j}_c$ is the macroscopic free-carriers current, $\rho_s(\mathbf{r})$ is the fluctuating charge density of the sluggish defects. One could take into account only electron mobility $\mu$ and diffusion coefficient $\kappa$, allowing for the defects sluggishness. Hereinafter $n<0$, $\mu<0$ and $\kappa>0$. As the equation of state, we use the Landau-Khalatnikov equation for the displacement z-component relaxation, but take into account the charged defects and correlation effects (i.e. $\partial^2 D_z / \partial \mathbf{r}^2 \neq 0$):

$$
\Gamma\frac{\partial D_z}{\partial t} + \alpha D_z + \beta D_z^3 - \lambda\frac{\partial^2 D_z}{\partial \mathbf{r}^2} = E_z
\tag{A.3}
$$

Parameter $\Gamma>0$ is the kinetic coefficient, $\alpha<0$, $\beta>0$, $\lambda>0$ are material parameters of the hypothetical pure (free of defects) sample.

The Debye screening radius $R_D = \sqrt{\kappa/4\pi n\mu} \approx \sqrt{\left| k_B T^* / 4\pi n e \right|}$ in accordance with Einstein relation $\mu/\kappa \approx e/k_B T^*$ (electron charge $e<0$) [33], [34]. Hereafter we suppose that homogeneous external field $E_0(t)$ is applied along the polar $z$ - axis. The film occupies the region $-l/2 \le z \le l/2$. We suppose that the potentials of electrodes are given, so that the inner field satisfies the condition:

$$\frac{1}{l}\int_{-l/2}^{l/2} E_z(\mathbf{r},t)\,dz = E_0(t)\,. \tag{A.4}$$

Hereinafter we introduce the representation

$$\mathbf{D}(\mathbf{r},t) = \mathbf{e}_z \overline{D(t)} + \delta\mathbf{D}(\mathbf{r},t)\,, \quad \mathbf{E}(\mathbf{r},t) = \mathbf{e}_z E_0(t) + \delta\mathbf{E}(\mathbf{r},t)\,,$$
$$\rho_S(\mathbf{r}) = \overline{\rho}_S + \delta\rho_S(\mathbf{r})\,, \quad n(\mathbf{r},t) = \overline{n} + \delta n(\mathbf{r},t)\,. \tag{A.5}$$

A dash designates the averaging of the function $f$ over the film volume $V$: $\overline{f}(t) = \frac{1}{V}\int_V f(\mathbf{r},t)d\mathbf{r}$. By definition $\overline{\delta f(\mathbf{r},t)} = 0$, where $f = \{n,\ \rho_S, E, D, \ldots\}$. Here $\mathbf{e}_z$ s the unit vector directed along z-axis, $\overline{E}$ is the applied uniform field $E_0(t)$ and $\overline{E}_{x,y} = 0$.

The spatial distribution of the deviations $\delta f(\mathbf{r},t)$ consist of the part caused by spontaneous displacement screening [1], [15], space-charged layers localized in the ultrathin screening regions near the electrodes/substrate [15], [16] and the one caused by microscopic modulation $\delta\rho_s$ with quasi-homogeneous spatial distribution. For the μm-thick sample the averaging over sample volume is equivalent to the statistical averaging with homogeneous distribution function. Using the distribution function properties under the averaging, we obtained from (A.1-3) the equations for the average quantities and their fluctuations:

$$\overline{n} = -\overline{\rho}_S \tag{A.6}$$

$$div(\delta\mathbf{D}) = 4\pi(\delta n + \delta\rho_S) \tag{A.7}$$

$$\mu\,\overline{n}\,E_0 + \mu\,\overline{\delta n\,\delta E_z} + \frac{\partial}{\partial t}\frac{\overline{D(t)}}{4\pi} = 0, \qquad \mu\,\overline{\delta n\,\delta E_{x,y}} = 0 \tag{A.8}$$

$$\mu\,\left(\delta n\,E_0\mathbf{e}_z + \overline{n}\delta\mathbf{E}\right) + \mu\left(\delta n\delta\mathbf{E} - \overline{\delta n\,\delta\mathbf{E}}\right) - \kappa\,\,grad\,\delta n + \frac{1}{4\pi}\frac{\partial}{\partial t}\delta\mathbf{D} = 0\,. \tag{A.9}$$

$$\Gamma\frac{\partial\overline{D}}{\partial t} + \left(\alpha + 3\beta\,\overline{\delta D^2}\right)\overline{D} + \beta\overline{D}^3 - \lambda\overline{\frac{\partial^2\delta D}{\partial\mathbf{r}^2}} = E_0(t), \tag{A.10}$$

$$\Gamma\frac{\partial}{\partial t}\delta D + \left(\alpha + 3\beta\,\overline{D}^2\right)\delta D + 3\beta\,\overline{D}\left(\delta D^2 - \overline{\delta D^2}\right) + \beta\delta D^3 - \lambda\frac{\partial^2\delta D}{\partial\mathbf{r}^2} = \delta E_z\,. \tag{A.11}$$

Note, that the absence of the space charge average density $0 = \overline{n} + \overline{\rho}_s$ in (A.6) means the sample electro-neutrality. The term $\mu\left(\delta n\,\delta\mathbf{E} - \overline{\delta n\,\delta\mathbf{E}}\right)$ in (A.8) can be interpreted as the fluctuating circular electric currents around charged defects, which does not contribute into the average macroscopic current. Also we denote $\delta D_z \equiv \delta D$ in (A.11).

The system of equations (A.6-11) is complete, because the quantities $\delta n$, $\delta \mathbf{E}$ can be expressed via the fluctuations of displacement $\delta D$ and $\delta \rho_s(\mathbf{r})$, namely in accordance with (A.7-9): $\delta n = \frac{1}{4\pi} div(\delta \mathbf{D}) - \delta \rho_S$,

$$\overline{\delta n \delta E_z} = -\overline{n} E_0 - \frac{\partial}{\partial t} \frac{\overline{D}(t)}{4\pi\mu} \text{ and } \delta E_z = -\frac{1}{4\pi\mu\overline{n}} \frac{\partial}{\partial t} \delta D - \frac{\delta n}{\overline{n}} E_0(t) + \frac{\kappa}{\mu\overline{n}} \frac{\partial}{\partial z} \delta n + \frac{\delta n \delta E_z - \overline{\delta n \delta E_z}}{\overline{n}}.$$

Then the spatial distribution and temporal evolution of the displacement $D(\mathbf{r},t)$ is determined by the non-linear system (A.10-11) allowing for the fact that (A.11) could be rewritten in the form:

$$\begin{aligned}
&\left(\Gamma + \frac{1}{4\pi\mu} \frac{1}{\overline{n}}\right)\frac{\partial}{\partial t}\delta D + \left(\alpha + 3\beta \overline{D}^2\right)\delta D + 3\beta \overline{D}\left(\delta D^2 - \overline{\delta D^2}\right) + \beta \delta D^3 - \\
&- \left(\lambda + \frac{\kappa}{4\pi\mu\overline{n}}\right)\frac{\partial^2 \delta D}{\partial \mathbf{r}^2} = -\frac{\kappa}{\mu\overline{n}}\frac{\partial \delta \rho_S}{\partial z} - \frac{E_0(t)}{\overline{n}}\delta n - \frac{\delta n \delta E_z - \overline{\delta n \delta E_z}}{\overline{n}}
\end{aligned}$$

(A.12)

However, in our opinion, only the adopted for the stochastic differential equations implicit numerical schemes can be used in order to obtain the numerical solutions of (A.6-11) supplemented by the appropriate distributions of fluctuations $\delta \rho_s(\mathbf{r})$ and boundary conditions for $\delta D(z = \pm l/2)$, which is beyond the scope of this paper. Similar problems for the ferroelectrics - ideal insulators were considered in details earlier (see e.g. [40], [41]). Therefore, hereinafter we consider only the average characteristics.

The coupled equations for $\overline{\delta D^2}$ and $\overline{\delta D \delta \rho_S}$ could obtained directly from (A.12) after multiplying over $\delta D$ or $\delta \rho_S$ and averaging. Along with (A.10) they have the form:

$$\Gamma \frac{\partial \overline{D}}{\partial t} + \left(\alpha + 3\beta \overline{\delta D^2}\right)\overline{D} + \beta \overline{D}^3 = \lambda \overline{\frac{\partial^2 \delta D}{\partial \mathbf{r}^2}} + E_0(t),$$

(A.13)

$$\begin{aligned}
&\left(\Gamma + \frac{1}{4\pi\mu} \frac{1}{\overline{n}}\right)\frac{\partial}{\partial t}\frac{\overline{\delta D^2}}{2} + \left(\alpha + 3\beta \overline{D}^2(t)\right)\overline{\delta D^2} + \beta \overline{\delta D^4} + 3\beta \overline{D}\,\overline{\delta D^3} = \\
&= \left(\lambda + \frac{\kappa}{4\pi\mu\overline{n}}\right)\overline{\delta D \frac{\partial^2 \delta D}{\partial \mathbf{r}^2}} - \frac{\kappa}{\mu\overline{n}}\overline{\delta D \frac{\partial \delta \rho_S}{\partial z}} - E_0 \overline{\frac{\delta D \delta n}{\overline{n}}} - \overline{\delta D \frac{\delta n \delta E_z - \overline{\delta n \delta E_z}}{\overline{n}}}
\end{aligned}$$

(A.14)

$$\begin{aligned}
&\left(\Gamma + \frac{1}{4\pi\mu} \frac{1}{\overline{n}}\right)\frac{\partial}{\partial t}\overline{\delta D \delta \rho_S} + \left(\alpha + 3\beta \overline{D}^2(t)\right)\overline{\delta D \delta \rho_S} + \beta \overline{\delta D^3 \delta \rho_S} + 3\beta \overline{D}\,\overline{\delta D^2 \delta \rho_S} = \\
&= \left(\lambda + \frac{\kappa}{4\pi\mu\overline{n}}\right)\overline{\delta \rho_S \frac{\partial^2 \delta D}{\partial \mathbf{r}^2}} - \frac{\kappa}{2\mu\overline{n}}\overline{\frac{\partial \delta \rho_S^2}{\partial z}} - E_0 \overline{\frac{\delta \rho_S \delta n}{\overline{n}}} - \overline{\delta \rho_S \frac{\delta n \delta E_z - \overline{\delta n \delta E_z}}{\overline{n}}}
\end{aligned}$$

(A.15)

Using the Bogolubov method [42], the fourth and the third power correlations in (A.14) and (A.15) could be approximated as:

$$\overline{\delta D^4} \approx \left(\overline{\delta D^2}\right)^2, \qquad \overline{\delta D^3} \approx 0, \qquad \overline{\delta D\left(\delta n\, \delta \mathbf{E} - \overline{\delta n\, \delta \mathbf{E}}\right)} \approx \overline{\delta D}\,\overline{\left(\delta n\, \delta \mathbf{E} - \overline{\delta n\, \delta \mathbf{E}}\right)} = 0$$

$$\overline{\delta D^3\, \delta\rho_S} \approx \overline{\delta D^2}\,\overline{\delta D\, \delta\rho_S}, \qquad \overline{\delta D^2\, \delta\rho_S} \approx 0, \qquad \overline{\delta\rho_S\left(\delta n\, \delta \mathbf{E} - \overline{\delta n\, \delta \mathbf{E}}\right)} \approx \overline{\delta\rho_S}\,\overline{\left(\delta n\, \delta \mathbf{E} - \overline{\delta n\, \delta \mathbf{E}}\right)} = 0 \tag{A.16}$$

The expressions for other correlations in (A.13-14) were derived in [29] for the bulk sample. In the present paper we approximate them for a finite film with thickness $l$ with the quasi-homogeneous distribution of charged defects (i.e. $\overline{\delta\rho_S^2} = const$, $\left(\overline{\delta\rho_S}\right)_{x,y} = 0$ for every $z$ from the region $-l/2 \leq z \leq l/2$).

In accordance with (A.7) built-in field $E_i \equiv \lambda\, \overline{\partial^2 \delta D / \partial \mathbf{r}^2}$ acquires the form:

$$E_i = \lambda\, \overline{\left(\frac{\partial \delta D}{\partial z}\right)}_{x,y} = \frac{4\pi\lambda}{l} \overline{\left(\delta n + \delta\rho_S\right)}_{x,y}\Big|_{-l/2}^{+l/2} \approx \frac{4\pi\lambda}{l} \overline{\left(\delta n\right)}_{x,y}\Big|_{-l/2}^{+l/2} \tag{A.17}$$

For a finite film $E_i$ is induced by the space charge layers accommodated near the non-equivalent boundaries $z = \pm l/2$ of examined heterostructure ($\overline{\delta n}(z \approx +l/2)_{x,y} \neq \overline{\delta n}(z \approx -l/2)_{x,y} \neq 0$).

Let us estimate the averaged gradient term correlations $\overline{\delta D\, \dfrac{\partial^2 \delta D}{\partial \mathbf{r}^2}}$ and $\overline{\delta\rho_S\, \dfrac{\partial^2 \delta D}{\partial \mathbf{r}^2}}$ using the Fourier representation of their complex amplitudes $\delta\widetilde{D}(\mathbf{r},t) = \mathrm{Re}\left[\delta D(\mathbf{r},t)\right]$ and $\delta\widetilde{\rho}_S(\mathbf{r},t) = \mathrm{Re}\left[\delta\rho_S(\mathbf{r},t)\right]$,

where $\delta\widetilde{D}(\mathbf{r},t) = \dfrac{1}{4\pi^2} \sum\limits_{m=1}^{\infty} \int\limits_{-\infty}^{\infty} dk_x dk_y\, C_m\left(k_x, k_y, t\right)\exp\left(i\left(k_x x + k_y y + \dfrac{2\pi m}{l} z\right)\right)$ and

$\delta\widetilde{\rho}_S(\mathbf{r}) = \dfrac{1}{4\pi^2} \sum\limits_{m=1}^{\infty} \int\limits_{-\infty}^{\infty} dk_x dk_y\, P_m\left(k_x, k_y\right)\exp\left(i\left(k_x x + k_y y + \dfrac{2\pi m}{l} z\right)\right)$. Using delta-function representation, we obtained that: $\left|\delta\widetilde{D}(\mathbf{r},t)\right|^2 = \sum\limits_{m=1}^{\infty} \int\limits_{-\infty}^{\infty} \dfrac{dk_x dk_y}{4\pi^2}\left|C_m\left(k_x, k_y, t\right)\right|^2$, and

$$\overline{\delta\widetilde{D}^*\, \frac{\partial^2 \delta\widetilde{D}}{\partial \mathbf{r}^2}} = -\sum\limits_{m=1}^{\infty} \int\limits_{-\infty}^{\infty} \frac{dk_x dk_y}{4\pi^2}\left|C_m\left(k_x, k_y, t\right)\right|^2 k_m^2\left(k_x, k_y\right), \qquad \overline{\delta\widetilde{\rho}_S^*\, \frac{\partial^2 \delta\widetilde{D}}{\partial \mathbf{r}^2}} = -\sum\limits_{m=1}^{\infty} \int\limits_{-\infty}^{\infty} \frac{dk_x dk_y}{4\pi^2} P_m^* C_m k_m^2\left(k_x, k_y\right)$$

here $k_m^2 = k_x^2 + k_y^2 + \dfrac{4\pi^2 m^2}{l^2}$. Under the conditions $d << l$ and $d << \lambda$, the fluctuation spatial spectrums $\left|C_m\left(k_x, k_y, t\right)\right|^2$ and $\left|P_m\left(k_x, k_y\right)\right|^2$ have well localized maximum at wave vector $\left|k_m\right| \approx \dfrac{2\pi}{d}$ (one dimensional case is depicted in Fig. 3). Thus in accordance with Laplace integration method and Lagrange theorem about the mean point one obtains that:

$$\overline{\delta D\, \frac{\partial^2 \delta D}{\partial \mathbf{r}^2}} \approx -\frac{1}{d^2}\, \overline{\delta D^2}, \qquad \overline{\delta\rho_S\, \frac{\partial^2 \delta D}{\partial \mathbf{r}^2}} \approx -\frac{1}{d^2}\, \overline{\delta\rho_S \delta D}. \tag{A.18}$$

When using integration over parts and equality $\delta n = \dfrac{1}{4\pi} div\left(\delta \mathbf{D}\right) - \delta\rho_S$, we transformed the terms

$$\frac{\kappa}{\mu \overline{n}}\overline{\left(\delta D \frac{\partial}{\partial z}\delta \rho_S\right)} = \frac{\kappa}{\mu \overline{n} l}\left(\overline{\delta D \,\delta \rho_S}\right)_{x,y}\Bigg|_{-l/2}^{+l/2} - \frac{4\pi\kappa}{\mu \overline{n}}\overline{\delta \rho_S\left(\delta \rho_S + \delta n\right)} \approx -\frac{4\pi\kappa}{\mu \overline{n}}\overline{\delta \rho_S\left(\delta \rho_S + \delta n\right)} \qquad \text{(A.19)}$$

$$-E_0(t)\frac{\overline{\delta D \delta n}}{\overline{n}} = E_0(t)\frac{\overline{\left(\delta \rho_S \delta D\right)}}{\overline{n}} - \frac{E_0(t)}{8\pi l\, \overline{n}}\left(\overline{\delta D^2}\right)_{x,y}\Bigg|_{-l/2}^{+l/2} \qquad \text{(A.20)}$$

Keeping in mind that the charged defects distribution is quasi-homogeneous (i.e. $\left(\overline{\delta \rho_S}\right)_{x,y} \approx 0$ for every $z$ from the region $-l/2 \le z \le l/2$), we derive that

$$\frac{\kappa}{\mu \overline{n} l}\left(\overline{\delta D \,\delta \rho_S}\right)_{x,y}\Bigg|_{-l/2}^{+l/2} \approx 0\,, \qquad \frac{\overline{\partial \,\delta \rho_S^2}}{\partial z} = \frac{1}{l}\left(\overline{\delta \rho_S^{\,2}}\right)_{x,y}\Bigg|_{-l/2}^{+l/2} \approx 0\,, \qquad \text{(A.21)}$$

$$\delta E_i = \frac{\left(\overline{\delta D^2}\right)_{x,y}}{8\pi l\, \overline{n}}\Bigg|_{-l/2}^{+l/2} = \frac{2\pi}{l\, \overline{n}}\overline{\left(\int\limits_{z_0}^{\overline{z}} dz(\delta n + \delta \rho_S)\right)^2}_{x,y}\Bigg|_{-l/2}^{+l/2} \sim \frac{\overline{(\delta n + \delta \rho_S)^2}}{l} \qquad \text{(A.22)}$$

The constant $z_0$ in (A.22) can be determined from the boundary conditions for $\overline{\delta D}_{x,y}$ at $z=0$.

Allowing for (A.16-22) three coupled equations (A.13-15) for the average displacement $\overline{D}$, its mean-square fluctuation $\overline{\delta D^2}$ and correlation $\overline{\delta D \,\delta \rho_S}$ acquire the form

$$\Gamma\frac{\partial \overline{D}}{\partial t} + \left(\alpha + 3\beta\,\overline{\delta D^2}\right)\overline{D} + \beta\,\overline{D}^3 = E_0(t) + E_i(l,t)\,, \qquad \text{(A.23)}$$

$$\frac{\Gamma_R}{2}\frac{\partial \overline{\delta D^2}}{\partial t} + \left(\alpha_R + 3\beta\overline{D}^2\right)\overline{\delta D^2} + \beta\left(\overline{\delta D^2}\right)^2 = E_0(t)\left(\frac{\overline{\delta D \,\delta \rho_S}}{\overline{n}} - \delta E_i\right) + \frac{4\pi k_B T}{\overline{n} e}\overline{\delta \rho_S\left(\delta \rho_S + \delta n\right)}\,, \quad \text{(A.24)}$$

$$\Gamma_R\frac{\partial \overline{\delta D \,\delta \rho_S}}{\partial t} + \left(\alpha_R + 3\beta\overline{D}^2 + \beta\,\overline{\delta D^2}\right)\overline{\delta D \,\delta \rho_S} = -E_0(t)\frac{\overline{\delta \rho_S \delta n}}{\overline{n}}\,. \qquad \text{(A.25)}$$

The coefficient $\alpha$ in (A.24-25) is renormalized by gradient terms as $\alpha_R = \alpha + \left(\dfrac{\lambda}{d^2} + \dfrac{\kappa}{4\pi\mu\overline{n}d^2}\right)$. The renormalization of Khalatnikov kinetic coefficient $\Gamma_R \equiv \Gamma + \tau_m$ in (A.24-25) is connected with the free carrier relaxation with characteristic time $\tau_m = \dfrac{1}{8\pi\mu\,\overline{n}}$.

In contract to the system (A.6-11), the system (A.23-25) is not complete, namely the correlation $\overline{\delta \rho_S \delta n}$, built-in field average value $E_i(l,t)$ and its fluctuation $\delta E_i(l,t)$ are unknown. Differential equations or exact expressions for these values should be obtained from (A.7-11). In the general case hypothetical complete exact system for average values and all their correlations could contain infinite number of differential equations and thus is not the simpler problem than the solution of equations (A.7-11). Allowing for made approximations (A.16) and (A.18) we have no special reason to derive an exact system. Instead we prefer to deal with much simpler approximate system and thus to estimate/postulate the correlation $\overline{\delta\rho_S\delta n}$ as it was demonstrated in [29] for a bulk sample. In [29] we derived that $-\overline{\delta\rho_S\delta n(t)} \approx \eta\,\overline{\delta\rho_S^2}$ ($0 \le \eta < 1$) at small enough external field amplitude. This result also could be applied to the film under external electric field small in comparison with microscopic ones, because each charged region $\delta\rho_S$ is surrounded by the electrons $\delta n$ in accordance with Debye screening mechanism (see Fig. 2b), but the smooth part of charge density $\delta n$ accommodated near the surface and related to the spontaneous displacement screening does not correlate with random charged defects distribution $\delta\rho_S$ (see solid curves in Fig. 3a).

The average pyroelectric coefficient $\overline{\gamma} = \partial\overline{D}/\partial T$ can be obtained by the differentiation of the coupled system (A.23-25) on the temperature $T$. We suppose that only the coefficient $\alpha = -\alpha_T(T_C - T)$ essentially depend on temperature in usual form: $\partial\alpha/\partial T = \alpha_T$ [1]. The temperature dependence of other coefficients $\beta$, $\lambda$ and $\Gamma$ could be neglected within the framework Landau-Ginsburg phenomenology for the second order phase transitions [1]. The temperature dependence $\partial\overline{n}/\partial T$ could be neglected at temperatures $T \gg E_a/k_B \sim 50\,K$ [33], [34]. Also we neglect the temperature dependences $\partial E_i/\partial T$ and $\partial\delta E_i/\partial T$ allowing for the same reasons ($E_i \sim \delta n$, $\delta E_i \sim \overline{(\delta n + \delta\rho_S)^2}$, $\overline{n} \sim \exp(-E_a/k_B T)$). In this way we obtained three coupled equations for pyroelectric coefficient $\overline{\gamma} = \partial\overline{D}/\partial T$, its correlation with displacement fluctuations $\overline{\delta\gamma\delta D} = \overline{\delta D\,\delta\delta D/\partial T}$ and charge defects density fluctuations $\overline{\delta\gamma\delta\rho_S} = \partial\overline{\delta D\,\delta\rho_S}/\partial T$:

$$\Gamma\frac{\partial\overline{\gamma}}{\partial t} + \left(\alpha + 3\beta\,\overline{\delta D^2} + 3\beta\overline{D}^2\right)\overline{\gamma} = -\left(\alpha_T + 3\beta\,\overline{\delta\gamma^2}\right)\overline{D}\,, \tag{A.29}$$

$$\Gamma_R\frac{\partial\overline{\delta\gamma\delta D}}{\partial t} + 2\left(\alpha_R + 2\beta\overline{\delta D^2} + 3\beta\overline{D}^2\right)\overline{\delta\gamma\delta D} = \begin{pmatrix} E_0(t)\dfrac{\overline{\delta\gamma\delta\rho_S}}{\overline{n}} - \left(\alpha_{RT} + 6\beta\,\overline{D}\,\overline{\gamma}\right)\overline{\delta D^2} + \\ + \dfrac{4\pi k_B}{\overline{n}e}\delta\rho_S\left(\delta\rho_S + \delta n\right) \end{pmatrix}, \tag{A.30}$$

$$\Gamma_R\frac{\partial\overline{\delta\gamma\delta\rho_S}}{\partial t} + \left(\alpha_R + \beta\overline{\delta D^2} + 3\beta\overline{D}^2\right)\overline{\delta\gamma\delta\rho_S} = -\left(\alpha_{RT} + 2\beta\overline{\delta\gamma\delta D} + 6\beta\overline{D}\,\overline{\gamma}\right)\overline{\delta D\delta\rho_S}\,. \tag{A.31}$$

The system determining the dielectric permittivity $\overline{\varepsilon} = \partial\overline{D}/\partial E_0$, its correlations $\overline{\delta\varepsilon\delta D} = \overline{\delta D\,\delta\delta D/\partial E_0}$ and $\overline{\delta\varepsilon\delta\rho_S} = \partial\overline{\delta D\,\delta\rho_S}/\partial E_0$ has the form:

$$\Gamma\frac{\partial\overline{\varepsilon}}{\partial t} + \left(\alpha + 3\beta\,\overline{\delta D^2} + 3\beta\overline{D}^2\right)\overline{\varepsilon} = 1 - 6\beta\,\overline{\delta\varepsilon\delta D}\,\overline{D}\,, \tag{A.32}$$

$$\Gamma_R \frac{\partial \overline{\delta\varepsilon\delta D}}{\partial t} + 2\left(\alpha_R + 2\beta\overline{\delta D^2} + 3\beta\overline{D}^2\right)\overline{\delta\varepsilon\delta D} = -\frac{\overline{\delta D\delta\rho_S}}{\overline{\rho}_s} - E_0(t)\frac{\overline{\delta\varepsilon\delta\rho_S}}{\overline{\rho}_s} - 6\beta\,\overline{\delta D^2}\,\overline{D}\,\overline{\varepsilon}, \quad \text{(A.33)}$$

$$\Gamma_R \frac{\partial \overline{\delta\varepsilon\delta\rho_S}}{\partial t} + \left(\alpha_R + \beta\overline{\delta D^2} + 3\beta\overline{D}^2\right)\overline{\delta\varepsilon\delta\rho_S} = \frac{\overline{\delta\rho_S\delta n}}{\overline{\rho}_S} - \beta\left(2\overline{\delta\varepsilon\delta D} + 6\overline{D}\,\overline{\varepsilon}\right)\overline{\delta D\delta\rho_S}. \quad \text{(A.34)}$$